\documentclass[twocolumn]{jpsj3}
\usepackage{txfonts}
\usepackage[top=1.in, bottom=1.3in, left=0.9in, right=0.3in]{geometry}

\newcommand{\CCI}{\ensuremath{\rm CeCoIn_5}}

\title{Terahertz Conductivity of the Heavy-Fermion State in CeCoIn$_5$}

\author{\name{Marc \surname{Scheffler}}$^1$\thanks{E-mail: scheffl@pi1.physik.uni-stuttgart.de}, \name{Thomas \surname{Weig}}$^1$, \name{Martin \surname{Dressel}}$^1$, \name{Hiroaki \surname{Shishido}}$^2$, \name{Yuta \surname{Mizukami}}$^2$, \name{Takahito \surname{Terashima}}$^3$, \name{Takasada \surname{Shibauchi}}$^2$ and \name{Yuji \surname{Matsuda}}$^2$}
\inst{$^1$1.\ Physikalisches Institut, Universit\"at Stuttgart, D-70550 Stuttgart, Germany \\
$^2$Department of Physics, Kyoto University, Kyoto 606-8502, Japan \\
$^3$Research Center for Low Temperature and Materials Science, Kyoto University, Kyoto 606-8502, Japan} 

\abst{The optical properties of thin films of the heavy-fermion compound \CCI, which were deposited by molecular beam epitaxy onto MgF$_2$ substrates, have been studied at frequencies 7 to 45~cm$^{-1}$ (corresponding to 0.2 to 1.3~THz) and temperatures 2 to 300~K. We observe an electrodynamic behavior which is typical for heavy fermions, namely Drude-like conductivity with a relaxation rate at rather low frequencies. This relaxation rate increases almost linearly with temperature up to at least 30~K. The coherent heavy-fermion state, characterized by an increase of the effective mass, continuously evolves upon cooling and is not fully developed for temperatures as low as 5~K.
}

\kword{CeCoIn$_5$, THz conductivity, optical conductivity, heavy fermion, thin film}

\begin{document}
\maketitle

Heavy-fermion materials are intermetallic compounds that serve as prime examples of strongly correlated electron systems. Among the plethora of heavy-fermion compounds, \CCI{} has taken a prominent role: its superconducting transition temperature $T_c$ of 2.3~K is the highest of the cerium-based heavy fermions,\cite{Petrovic2001} and its phase diagram features a candidate for a Fulde-Ferrell-Larkin-Ovchinnikov state \cite{Matsuda2007} as well as quantum-critical and Fermi-liquid phases.\cite{Paglione2003,Bianchi2003} Despite numerous experimental approaches, no complete understanding is reached concerning the electronic properties of \CCI. Optical spectroscopy addresses charge dynamics as well as electronic and magnetic excitations, and thus is an appropriate technique to explore many of the interesting aspects of \CCI. So far, optical studies in the infrared frequency range covered temperatures down to 8~K,\cite{Singley2002,Mena2005,Burch2007} whereas the only electrodynamic measurements at lower frequencies, in the THz and GHz ranges, exclusively focused on the superconducting state.\cite{Ormeno2002,Nevirkovets2008,SudhakarRao2009,Truncik2012}
 
The reason why there is no detailed THz study of \CCI{} yet is that the reflectivity of bulk \CCI{} in this frequency range is too close to unity to be measured accurately. The only viable experimental solution to address heavy-fermion metals at these frequencies is measuring the transmission of thin film samples,\cite{Dressel2002c,Ostertag2010,Ostertag2011,Bosse2012} but so far no suitable thin films of \CCI{} were available. With our recent improvements in the thin film growth of the CeIn$_3$ and CeCoIn$_5$ material systems,\cite{Shishido2010, Mizukami2011} precise THz measurements in transmission geometry now become feasible.

At the present state, there are several open questions concerning the low-energy electrodynamics of heavy fermions. The first addresses the Drude relaxation:\cite{Dressel2002a} the relaxation rate of heavy fermions is expected at frequencies as low as a few GHz.\cite{Millis1987,Scheffler2013} Experimentally, this was confirmed for some uranium-based heavy fermions,\cite{Tran2002,Scheffler2005c,Scheffler2010,Scheffler2006} whereas there is only little comparable microwave data available concerning the Drude relaxation of cerium-based heavy fermions.\cite{Beyermann1988a,Awasthi1993} The second open question concerns possible low-energy excitations that have been found in the THz conductivity of some uranium-based heavy fermions,\cite{Donovan1997,Dressel2002c,Dressel2002b,Ostertag2011} but not yet for cerium-based systems.\cite{Bosse2012,Beyermann1988a,Awasthi1993} Additional THz studies on cerium-based heavy fermions might indicate whether this is a fundamental difference between uranium- and cerium-based materials. Optical measurements can also address more generic questions concerning charge carriers in metals, and we will show how our THz studies might shed new light on the unconventional properties of \CCI{} at intermediate temperatures, where \CCI{} is considered a non-Fermi liquid (NFL) material that might be governed by the vicinity of a quantum-critical point.\cite{Paglione2003,Bianchi2003,Aynajian2012}

\begin{figure}
\centering
    \includegraphics[width=8.4cm]{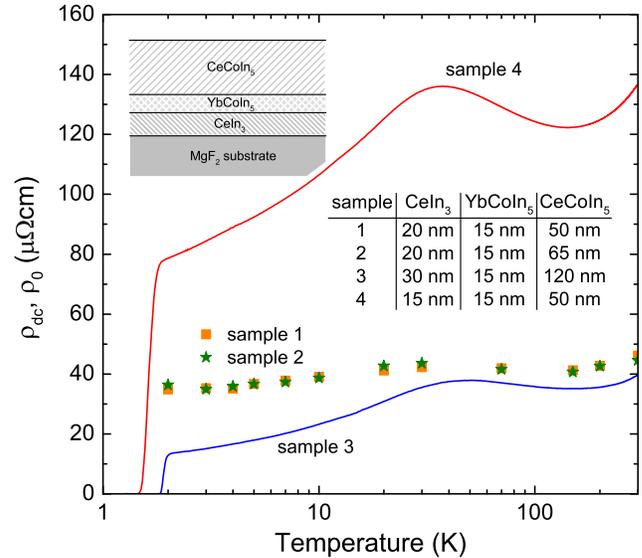}
\caption{\label{FigRdc} (Color online) Temperature dependence of dc resistivity $\rho_{\textrm dc}$ of thin films (samples 3 and 4) similar to those (samples 1 and 2) studied with THz spectroscopy. For samples 1 and 2, $\rho_0 = 1/\sigma_0$ determined from Drude fits to the THz data is shown for comparison. The upper left inset schematically shows the layers of the thin film samples. The thicknesses of the different layers for the four samples are listed in the inset table.}
\end{figure}

For our THz studies, thin films of \CCI{} with sufficiently high transmission are a demanding prerequisite. Despite the efforts of several groups during the last decade,\cite{Soroka2007, Izaki2007, Zaitsev2009, Haenisch2010} growing high-quality \CCI{} thin films remained challenging. For the present experiment, we have optimized the growth of \CCI{} thin films on MgF$_2$(001) substrates with molecular beam epitaxy.\cite{Shishido2010, Mizukami2011} The major difficulty here is that the quality of \CCI{} films typically improves with increasing film thickness, but for THz transmission studies the thickness of a metallic film should stay below 100~nm. We have found that the growth of \CCI{} on MgF$_2$ can be improved by CeIn$_3$ and YbCoIn$_5$ buffer layers, as shown schematically in the inset of Fig.\ref{FigRdc}.\cite{Mizukami2011,Shimozawa2012} Using these metallic buffer layers has the disadvantage that the overall thin-film sample now has three conducting layers and that any experiment probes a combination of these three layers, whereas our interest here is focused exclusively on the \CCI{} layer. Fortunately, as we will show, the physical response of our samples is dominated by \CCI{} despite the presence of the buffer layers. The thicknesses of the different metallic layers for the four samples discussed in this work are listed as inset to Fig.\ \ref{FigRdc}.

Direct evidence for the dominant role of \CCI{} in our samples comes from dc measurements. Fig.\ \ref{FigRdc} shows the temperature dependence of the effective dc resistivity $\rho_{\textrm dc}$ of samples 3 and 4, which have different layer thicknesses than samples 1 and 2 studied with THz spectroscopy, but otherwise were prepared in the same way.\cite{CommentDegradationRdcTHz} The dc resistivity values were determined by simply multiplying the sheet resistance (from the standard four-probe measurements) with the total thickness of the metallic trilayer. This procedure obliterates the three different layers, but in our case is justified by the experimental data, since we cannot deduce any contribution of the CeIn$_3$ and YbCoIn$_5$ layers. The temperature dependence in Fig.\ \ref{FigRdc} shows the well-known characteristics of \CCI:\cite{Petrovic2001,Singley2002} upon cooling from room temperature, the resistivity first decreases slightly and then increases below 140~K due to incoherent Kondo scattering. Passing a resistivity maximum around 40~K, the material enters the coherent heavy-fermion state with considerably decreasing resistivity and finally develops superconductivity below 2~K.
While these fingerprints of \CCI{} are evident in our samples, there are none of the characteristics of the other two layers: CeIn$_3$ is a heavy-fermion metal with an antiferromagnetic transition around 10~K (for thin films) that features as a pronounced kink in the temperature-dependent resistivity,\cite{Shishido2010,Foyevtsov2009,Zaitsev2012} which is absent in our data. YbCoIn$_5$ is a conventional metal with monotonic temperature dependence of the resistivity.\cite{Mizukami2011} From the dominant \CCI{} features in the dc resistivity we conclude that our description of the electronic properties of our trilayer simply in terms of \CCI{} is a valid approximation.

We performed THz transmission measurements on the \CCI{} samples as well as a bare MgF$_2$ reference substrate using a frequency-domain THz spectrometer, which is based on several backward wave oscillators as coherent, monochromatic and frequency-tunable THz sources and a Golay cell or a bolometer as radiation detectors.\cite{Gorshunov2008} We covered frequencies between 7~cm$^{-1}$ and 45~cm$^{-1}$ and temperatures from 2~K to 300~K using a home-built optical cryostat.\cite{Ostertag2011}
One challenge of these measurements is that the \CCI{} thin films degrade when exposed to air. Therefore, the freshly-grown films were shipped from Kyoto to Stuttgart in evacuated glass tubes, and exposure to air was minimized (time between opening of glass tube and inserting samples into the optical cryostat was less than 30 minutes). With the samples in the helium gas inside the cryostat, we observed degradation of the samples on a time scale of weeks (by comparing the room temperature transmission). The data presented in the following were obtained within 60 hours, and within our experimental sensitivity we did not observe any degradation during this period.

\begin{figure}
\centering
    \includegraphics[width=8.0cm]{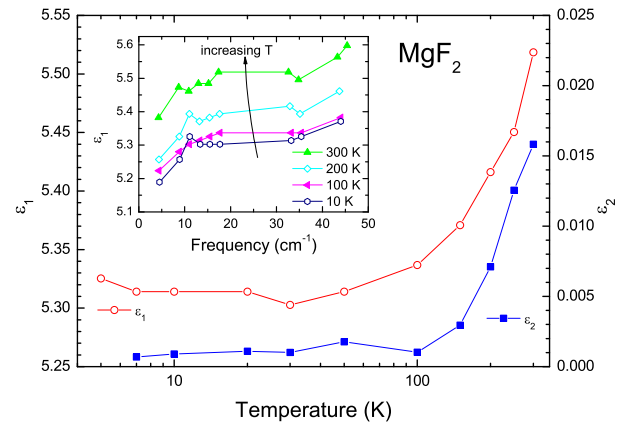}
\caption{\label{FigMgF2} (Color online) Temperature dependence of the dielectric constant $\epsilon = \epsilon_1 + i \epsilon_2$ of the MgF$_2$ substrate at frequency 33~cm$^{-1}$. The inset shows the frequency dependence of $\epsilon_1$ for exemplary temperatures.}
\end{figure}

The transmission data of the bare MgF$_2$ substrate were analyzed by fitting the Fabry-Perot maxima of this dielectric slab with the real and imaginary parts of the dielectric constant $\epsilon = \epsilon_1 + i \epsilon_2$ as fitting parameters.\cite{Dressel2002a} The results of these fits are displayed in Fig.\ \ref{FigMgF2}. These data for $\epsilon$ of MgF$_2$ are then used as an input for the analysis of the Fabry-Perot maxima of the \CCI{} samples. We model the complete sample as a two-layer system: MgF$_2$ as the first layer and the metallic films as the second layer. Here we describe the triple metal layer CeIn$_3$/YbCoIn$_5$/CeCoIn$_5$ as a single effective layer and attribute its property to \CCI. This simplification is justified by the dc measurements described above, but also by the outcome of the THz measurements, as discussed below.

\begin{figure}
\centering
    \includegraphics[width=8.4cm]{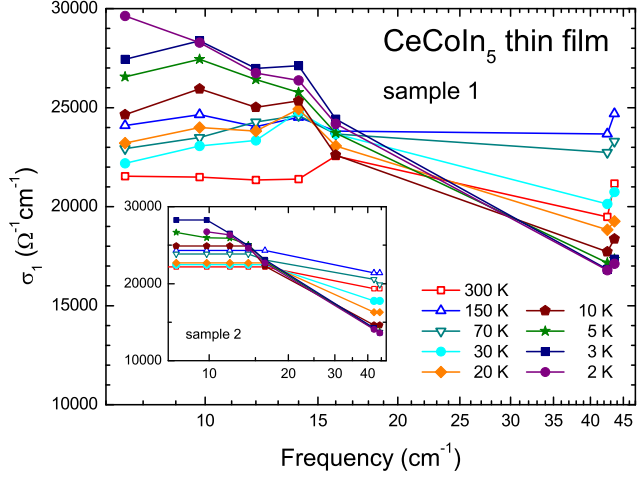}
\caption{\label{FigSpectra} (Color online) Conductivity spectra of 
two \CCI{} samples for a set of different temperatures.}
\end{figure}

By fitting the Fabry-Perot maxima in the transmission data of the thin-film samples to well-established formulas,\cite{Dressel2002a} we directly obtain the real part $\sigma_1$ of the optical conductivity of the film, without the need of a Kramers-Kronig transformation. The resulting $\sigma_1$ data of the thin films are displayed in Fig.\ \ref{FigSpectra}. For temperatures of 70~K and above, the conductivity spectra are basically flat, which is consistent with conventional metallic behavior with a Drude relaxation rate at much higher frequencies, i.e.\ in the infrared range. For temperatures of 30~K and below, the spectra reveal a decrease toward higher frequencies, and this roll-off becomes more pronounced with decreasing temperature. This behavior is explained by the development of the coherent heavy-fermion state which goes hand in hand with a strong decrease of the Drude relaxation rate.
The temperature where this effect starts, around 50~K, is consistent with previous studies on \CCI{} using other techniques.\cite{Petrovic2001,Aynajian2012} Our absolute values of the optical conductivity are somewhat lower than what is inferred from previous optical studies at higher frequencies,\cite{Singley2002,Mena2005} but this is due to the fact that they measured single crystals which typically have lower scattering rates than thin films. 

\begin{figure}
\centering
    \includegraphics[width=8.0cm]{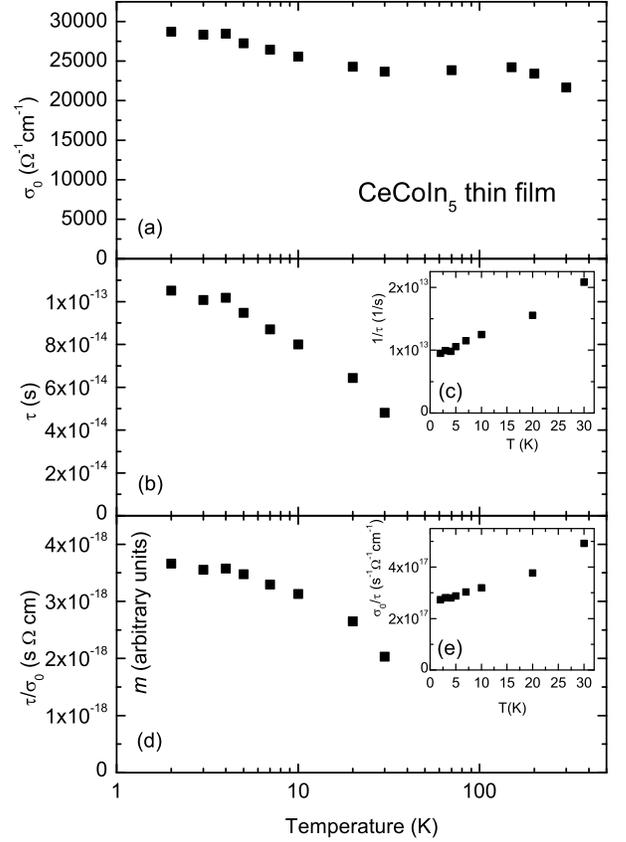}
\caption{\label{FigSigma0tau} Temperature dependence of (a) dc conductivity $\sigma_0$ and (b) relaxation time $\tau$ as determined from Drude fits to the conductivity spectra of \CCI. Their (d) ratio $\tau / \sigma_0$ is a measure for the effective mass $m$. The insets (c) and (e) display a roughly linear temperature dependence of relaxation rate $1/\tau$ and inverse effective mass $1/m \propto \sigma_0/\tau$, respectively.}
\end{figure}

For further discussion of our data, we have fitted the conductivity spectra with the simple Drude formula $\sigma_1=\sigma_0 /(1+\omega^2 \tau^2)$ where $\sigma_0$ is the dc conductivity, $\tau$ the Drude relaxation time, and $\omega$ the angular frequency of the radiation. The evolution of the Drude parameters $\sigma_0$ and $\tau$ as a function of temperature is plotted in Fig.\ \ref{FigSigma0tau}~(a) and (b), respectively. 
The values of $\sigma_0$ are very close to the values of $\sigma_1$ at the lowest measured frequencies in Fig.\ \ref{FigSpectra}. This might indicate that the optical conductivity is basically constant at lower frequencies, although additional features cannot be ruled out.\cite{Dressel2002b,Ostertag2011} For comparison with the dc resistivity measured for samples 3 and 4, in Fig.\ \ref{FigRdc} we have plotted $\rho_0 = 1/\sigma_0$ obtained from the Drude fits for samples 1 and 2. Again we find the characteristic features for \CCI, i.e.\ upon cooling from room temperature a decrease in $\rho_0$ with minimum around 150~K, then a slight increase with maximum between 70~K and 30~K, and a decrease for lower temperatures. This last decrease for temperatures below 30~K is considerably weaker than that of $\rho_{\textrm dc}$ for samples 3 and 4; this we attribute to sample degradation of samples 1 and 2 after the film growth and before the THz measurements.
But the overall temperature dependence of $\rho_0$ from THz data substantiates our assumption that the optical response of our samples is governed by the \CCI{} layer and that for our analysis we can neglect the buffer layers.
From the dc measurements on samples 3 and 4 we also know that the temperature where superconductivity sets in for our thin films is below 2~K, i.e.\ not accessible within this THz study.

Now we return to the outcome of the Drude fits in Fig.\ \ref{FigSigma0tau}. In general, one expects $\sigma_0(T)$ and $\tau(T)$ of a metal to scale if effective mass and charge carrier density remain constant. But for our heavy-fermion material the effective mass strongly depends on temperature, and therefore $\sigma_0$ and $\tau$ will not scale over the complete range of temperature. In fact, our data clearly indicate that above 5~K the coherent heavy-fermion state evolves and is fully developed only at lower temperatures: in Fig.\ \ref{FigSigma0tau}~(d), we show the ratio $\tau / \sigma_0$, which indicates the temperature dependence of the effective mass $m$, as within the Drude picture $\sigma_0 = n e^2 \tau / m$ is expected, and therefore $\tau / \sigma_0 = m / (n e^2) \propto m$ if the effective charge carrier density $n$ is 
constant ($e$ is the electron charge). In our specific case, the dc conductivity $\sigma_0$ does not change much within the studied temperature range, and therefore the temperature dependence of $m$ basically
follows the temperature-dependent $\tau$ as determined from our THz measurements.
I.e.\ the almost constant $\sigma_0$ indicates that the mean free path does not depend much on temperature, and in this case the evolutions of $m$ and $\tau$ are expected to scale upon the transition from
conventional metal at high temperature to
heavy-fermion state at low temperature.\cite{Millis1987,Scheffler2013}
As evident from Fig.\ \ref{FigSigma0tau}~(d), the effective mass increases strongly upon cooling, at least in the temperature range where we can clearly assign a relaxation rate from the conductivity spectra, which is below 30~K for our measurements.

Below 5~K, the temperature dependence of the effective mass levels off. Such a behavior is frequently found for heavy-fermion systems, i.e.\ an increase of the effective mass upon cooling for temperatures of the order of 10~K and reaching a constant value for temperatures of the order of 1~K. Nevertheless a precise determination of the effective mass is difficult in this temperature range for many experimental techniques that are the prime measurements of the effective mass at very low temperatures, such as the specific heat or the de Haas-van Alphen effect.\cite{Petrovic2001,Kim2001,McCollam2005} In this regard, measurements of the optical conductivity such as ours offer access to the temperature evolution of the effective mass also at intermediate temperatures. Our observation of a pronounced effective mass increase upon cooling around 30~K is consistent with a recent scanning tunneling microscopy (STM) study that uses quasiparticle interference to address this emergence of the heavy charge carriers,\cite{Aynajian2012} but the weak temperature dependence of the effective mass below 10~K is at odds with what was concluded from specific heat measurements.\cite{Kim2001} In the same STM work, the lifetime of the heavy-fermion state was studied and a linear temperature dependence of the inverse lifetime was found. We can compare this to the relaxation rate $1/\tau$ determined from our optical experiments, and we also find a roughly linear temperature dependence as shown in Fig.\ \ref{FigSigma0tau}~(c). But as can be concluded from Fig.\ \ref{FigSigma0tau}~(e), also the inverse effective mass has roughly linear temperature dependence, which, via the direct relation of effective mass and relaxation time,\cite{Millis1987} leads to the linear-in-$T$ behavior of the relaxation rate. In other words this latter linear temperature dependence is not related to the scattering processes but to the mass enhancement. This explanation is in contrast to the recent STM study, where the linear temperature dependence of the inverse lifetime was interpreted as a possible indication of quantum criticality.\cite{Aynajian2012}

Finally, we discuss our THz data in the context of Drude relaxation and low-frequency excitations of heavy fermions.
Our low-temperature data display an increase in $\sigma_1$ with decreasing frequency, consistent with simple Drude behavior, i.e.\ there is no suppression of the conductivity toward lower frequencies that might signal a low-frequency excitation as observed in certain uranium-based heavy fermions.\cite{Donovan1997,Dressel2002c,Dressel2002b,Ostertag2011} As a consequence, we also do not have any indications for an extremely slow Drude relaxation at GHz frequencies.\cite{Scheffler2005c} Instead, we find Drude relaxation in the THz frequency range, which for a thin film of a normal metal would be a very low frequency, but for a heavy-fermion material is in line with previous studies on other cerium compounds.\cite{Webb1986,Beyermann1988a,Awasthi1993,Bosse2012}

Future studies should extend the spectral range of the present work to even lower frequencies, including GHz frequencies. Ideally, these should be performed on \CCI{} thin films of yet higher quality and without the metallic buffer layers; ongoing progress in thin film growth
might make this possible.\cite{Shimozawa2012}
Such experiments should conclusively show the presence or absence of a low-frequency excitation at low THz/high GHz frequencies. Recent advances in low-temperature microwave spectroscopy on superconductors should even allow broadband 
access to the electrodynamics of the superconducting state of \CCI,\cite{Scheffler2013,Steinberg2008,Liu2011,Steinberg2012} thus adding the full frequency dependence to the previous studies of the temperature dependence below $T_c$ at fixed frequencies.\cite{Ormeno2002,Nevirkovets2008,Truncik2012}

\begin{acknowledgments}

We thank Stefan Kirchner and Ali Yazdani for helpful discussions.
We acknowledge financial support by the DFG. The work in Kyoto was supported by Grant-in-Aid for the Global COE program ``The Next Generation of Physics, Spun from Universality and Emergence'', Grant-in-Aid for Scientific Research on Innovative Areas ``Heavy Electrons'' from MEXT, and KAKENHI from JSPS.

\end{acknowledgments}






\begin{thebibliography}{99}

\bibitem{Petrovic2001}C. Petrovic, P. G. Pagliuso, M. F. Hundley, R. Movshovich, J. L. Sarrao, J. D. Thompson, Z. Fisk, and P. Monthoux:
J. Phys.: Condens. Matter \textbf{13} (2001) L337.


\bibitem{Matsuda2007}Y. Matsuda and H. Shimahara:
J. Phys. Soc. Jpn. \textbf{76} (2007) 051005.

\bibitem{Paglione2003}J. Paglione, M. A. Tanatar, D. G. Hawthorn, E. Boaknin, R. W. Hill, F. Ronning, M. Sutherland, L. Taillefer, C. Petrovic and P. C. Canfield:
Phys. Rev. Lett. \textbf{91} (2003) 246405.

\bibitem{Bianchi2003}A. Bianchi, R. Movshovich, I. Vekhter, P. G. Pagliuso, and J. L. Sarrao:
Phys. Rev. Lett. \textbf{91} (2003) 257001.


\bibitem{Singley2002}E. J. Singley, D. N. Basov, E. D. Bauer, and M. B. Maple:
Phys. Rev. B \textbf{65} (2002) 161101(R).

\bibitem{Mena2005}F. P. Mena, D. van der Marel, and J. L. Sarrao:
Phys. Rev. B \textbf{72} (2005) 045119.

\bibitem{Burch2007}K. S. Burch, S. V. Dordevic, F. P. Mena, A. B. Kuzmenko, D. van der Marel, J. L. Sarrao, J. R. Jeffries, E. D. Bauer, M. B. Maple, and D. N. Basov:
Phys. Rev. B \textbf{75} (2007) 054523.

\bibitem{Ormeno2002}R. J. Ormeno, A. Sibley, C. E. Gough, S. Sebastian, and I. R. Fisher:
Phys. Rev. Lett. \textbf{88} (2002) 047005.

\bibitem{Nevirkovets2008}I. P. Nevirkovets, O. Chernyashevskyy, C. Petrovic, J. B. Ketterson, and B. K. Sarma:
Physica C \textbf{468} (2008) 432.

\bibitem{SudhakarRao2009}G. V. Sudhakar Rao, S. Ocadlik, M. Reedyk, and C. Petrovic:
Phys. Rev. B \textbf{80} (2009) 064512.

\bibitem{Truncik2012}C. J. S. Truncik, W. A. Huttema, P. J. Turner, S. \"Ozcan, N. C. Murphy, P. R. Carrière, E. Thewalt, K. J. Morse, A. J. Koenig, J. L. Sarrao, and D. M. Broun:
arXiv:1210.5571v1.

\bibitem{Dressel2002c}M. Dressel, N. Kasper, K. Petukhov, D.N. Peligrad, B. Gorshunov, M. Jourdan, M. Huth, and H. Adrian:
Phys. Rev. B \textbf{66} (2002) 035110.

\bibitem{Ostertag2010}J. P. Ostertag, M. Scheffler, M. Dressel, and M. Jourdan:
Phys. Status Solidi B \textbf{247} (2010) 760.

\bibitem{Ostertag2011}J. P. Ostertag, M. Scheffler, M. Dressel, and M. Jourdan:
Phys. Rev. B \textbf{84} (2011) 035132.

\bibitem{Bosse2012}G. Boss\'e, L. S. Bilbro, R. Vald\'es Aguilar, L. Pan, W. Liu, A. V. Stier, Y. Li, L. H. Greene, J. Eckstein, and N. P. Armitage:
Phys. Rev. B \textbf{85} (2012) 155105.


\bibitem{Shishido2010}H. Shishido, T. Shibauchi, K. Yasu, T. Kato, H. Kontani, T. Terashima, and Y. Matsuda:
Science \textbf{327} (2010) 980.

\bibitem{Mizukami2011}Y. Mizukami, H. Shishido, T. Shibauchi, M. Shimozawa, S. Yasumoto, D. Watanabe, M. Yamashita, H. Ikeda, T. Terashima, H. Kontani, and Y. Matsuda:
Nature Physics \textbf{7} (2011) 849.


\bibitem{Dressel2002a}M. Dressel and G. Gr\"{u}ner:
{\it Electrodynamics of Solids} (Cambridge University Press, Cambridge, 2002).

\bibitem{Millis1987}A. J. Millis and P. A. Lee:
Phys. Rev. B \textbf{35} (1987) 3394.

\bibitem{Scheffler2013}M. Scheffler, K. Schlegel, C. Clauss, D. Hafner, C. Fella, M. Dressel, M. Jourdan, J. Sichelschmidt, C. Krellner, C. Geibel, F. Steglich:
Phys. Status Solidi B \textbf{250}, 439 (2013).


\bibitem{Tran2002}P. Tran, S. Donovan, and G. Gr\"uner:
Phys. Rev. B \textbf{65} (2002) 205102.

\bibitem{Scheffler2005c}M. Scheffler, M. Dressel, M. Jourdan, and H. Adrian:
Nature \textbf{438} (2005) 1135.

\bibitem{Scheffler2010}M. Scheffler, M. Dressel, and M. Jourdan:
Eur. Phys. J. B \textbf{740} (2010) 331.

\bibitem{Scheffler2006}M. Scheffler, M. Dressel, M. Jourdan,  and H. Adrian:
Physica B \textbf{378-380} (2006) 993.


\bibitem{Beyermann1988a}W. P. Beyermann, G. Gr\"uner, Y. Dalichaouch, and M. B. Maple:
Phys. Rev. Lett. \textbf{60} (1988) 216.

\bibitem{Awasthi1993}A. M. Awasthi, L. Degiorgi, G. Gr\"uner, Y. Dalichaouch, and M. B. Maple:
Phys. Rev. B \textbf{48} (1993) 10692.


\bibitem{Donovan1997}S. Donovan, A. Schwartz, and G. Gr\"uner:
Phys. Rev. Lett. \textbf{79} (1997) 1401.

\bibitem{Dressel2002b}M. Dressel, N. Kasper, K. Petukhov, B. Gorshunov, G. Gr\"uner, M. Huth, and H. Adrian:
Phys. Rev. Lett. \textbf{88} (2002) 186404.


\bibitem{Aynajian2012}P. Aynajian, E. H. da Silva Neto, A. Gyenis, R. E. Baumbach, J. D. Thompson, Z. Fisk, E. D. Bauer, and A. Yazdani:
Nature \textbf{486} (2012) 201.


\bibitem{Soroka2007}O. K. Soroka, G. Blendin, and M. Huth:
J. Phys.: Condens. Matter \textbf{19} (2007) 056006.

\bibitem{Izaki2007}M. Izaki, H. Shishido, T. Kato, T. Shibauchi, Y. Matsuda, and T. Terashima:
Appl. Phys. Lett. \textbf{91} (2007) 122507.

\bibitem{Zaitsev2009}A. G. Zaitsev, A. Beck, R. Schneider, R. Fromknecht, D. Fuchs, J. Geerk, H. v. L\"ohneysen:
Physica C \textbf{469} (2009) 52.

\bibitem{Haenisch2010}J. H\"anisch, F. Ronning, R. Movshovich, V. Matias:
Physica C \textbf{470} (2010) S568.

\bibitem{Shimozawa2012}M. Shimozawa, T. Watashige, S. Yasumoto, Y. Mizukami, M. Nakamura, H. Shishido, S. K. Goh, T. Terashima, T. Shibauchi, and Y. Matsuda:
Phys. Rev. B \textbf{86}, 144526 (2012).

\bibitem{CommentDegradationRdcTHz}The fast sample degradation has prohibited dc and THz measurements to be performed on the same sample.

\bibitem{Foyevtsov2009}O. Foyevtsov, and M. Huth:
J. Phys.: Conf. Ser. \textbf{150} (2009) 052057.

\bibitem{Zaitsev2012}A. G. Zaitsev, A. Beck, D. Fuchs, R. Fromknecht, M. Wissinger, R. Schneider, J. Geerk, H. v. L\"ohneysen:
J. Low Temp. Phys. \textbf{168} (2012) 90.


\bibitem{Gorshunov2008}B. P. Gorshunov, A.A. Volkov, A. S. Prokhorov, and I.E. Spektor:
Physics of the solid state \textbf{50} (2008) 2001.


\bibitem{Kim2001} J. S. Kim, J. Alwood, G. R. Stewart, J. L. Sarrao, and J. D. Thompson:
Phys. Rev. B \textbf{64} (2001) 134524.

\bibitem{McCollam2005}A. McCollam, S. R. Julian, P. M. C. Rourke, D. Aoki, and J. Flouquet:
Phys. Rev. Lett. \textbf{94} (2005) 186401.


\bibitem{Webb1986}B. C. Webb, A. J. Sievers, and T. Mihalisin:
Phys. Rev. Lett. \textbf{57} (1986) 1951.


\bibitem{Steinberg2008}K. Steinberg, M. Scheffler, and M. Dressel:
Phys. Rev. B \textbf{77} (2008) 214517.

\bibitem{Liu2011}W. Liu, M. Kim, G. Sambandamurthy, and N. P. Armitage:
Phys. Rev. B \textbf{84} (2011) 024511.

\bibitem{Steinberg2012}K. Steinberg, M. Scheffler, and M. Dressel:
Rev. Sci. Instrum. \textbf{83} (2012) 024704.


\end{thebibliography}
\end{document}